\documentclass[fleqn,twoside]{article}
\usepackage[headings]{espcrc2}

\readRCS
$Id: espcrc2.tex,v 1.2 2004/02/24 11:22:11 spepping Exp $
\usepackage{graphicx}

\newcommand{\fs}{\footnotesize}
\newcommand{\N}{\nonumber}
\newcommand{\eps}{\varepsilon}
\newcommand{\ep}{\varepsilon}

\newcommand{\gsim}{\raisebox{-0.07cm   }
{$\, \stackrel{>}{{\scriptstyle\sim}}\, $}}

\newcommand{\AmS}{{\protect\the\textfont2
  A\kern-.1667em\lower.5ex\hbox{M}\kern-.125emS}}

\title{\vspace*{-10mm}
{\footnotesize DESY 08--082; SFB-/CPP-08-38 \hfill 
}\\
First $O(\alpha_s^3)$ heavy flavor contributions to
       deeply inelastic scattering
       \thanks{Presented at Loops and Legs in
               Quantum Field Theory, Sondershausen, Germany, 2008. This paper 
was supported in part by SFB-TR-9:
                 Computergest\"utze Theoretische Teilchenphysik, and
                 Studienstiftung des Deutschen Volkes.
}%
        }
\author{I. Bierenbaum\address[DESY]{Deutsches Elektronen-Synchrotron, DESY,
                     Platanenallee 6, D-15738 Zeuthen, Germany},
        J. Bl\"umlein\addressmark[DESY]
        and S. Klein\addressmark[DESY] }

\runtitle{  }
\runauthor{ }
\begin{document}
\begin{abstract}
\noindent  
  In the asymptotic limit $Q^2 \gg m^2$, the heavy flavor Wilson coefficients 
  for deep--inelastic scattering factorize into the massless Wilson coefficients 
  and the universal heavy flavor operator matrix elements resulting from  
  light--cone expansion. In this way, one can calculate all but the power 
  corrections in $(m^2/Q^2)^k, k > 0$. The heavy flavor operator matrix elements 
  are known to ${\sf NLO}$. We present the last $2$--loop result missing in the 
  unpolarized case for the renormalization at $3$--loops and first 3--loop 
  results for terms proportional to the color factor $T_F^2$ in Mellin--space. 
  In this calculation, the corresponding parts of the ${\sf NNLO}$ anomalous 
  dimensions \cite{LARIN,MVVandim} are obtained as well.
\vspace{1pc}
\end{abstract}
%
\maketitle
%
%
\section{Introduction}

\noindent
The unpolarized deep--inelastic double differential scattering cross-section
can be expressed in terms of the structure functions $F_2(x,Q^2)$ and $F_L(x,Q^2)$ 
in case of single photon exchange. In the small $x$ region, both structure 
functions 
contain large $c\overline{c}$--contributions of the order of 20-40~\%. 
Precision
extractions of parton distribution functions and the measurement of $\Lambda_{\rm QCD}$
therefore require to extend the description of these contributions to the 
$O(\alpha_s^3)$ terms as reached in the massless case. The complete NLO corrections
were calculated semi--analytically  in $x$--space, \cite{HEAV1}, for 
which a 
fast implementation in Mellin $N$--space was given in \cite{HEAV2}.  For 
$Q^2\: \gsim 
\:10\:m_c^2$, one observes that $F_2^{c\bar{c}}(x,Q^2)$ is very well
described by its asymptotic expression in the limit ${Q^2\gg m^2}$, \cite{BUZA}.
In this kinematic range, one can calculate the heavy flavor Wilson 
coefficients, the 
perturbative part of the structure functions $F^{c\bar{c}}_2(x,Q^2)$ and  
$F^{c\bar{c}}_L(x,Q^2)$, analytically, which has been done for 
$F^{c\bar{c}}_2(x,Q^2)$ 
to 2--loop order in \cite{BUZA,BBK1} and for $F^{c\bar{c}}_L(x,Q^2)$
to 3--loop order in \cite{BFNK}. First steps towards an asymptotic 
3--loop calculation 
for $F_2^{c\bar{c}}(x,Q^2)$ have been made by the present authors by 
calculating  the O($\eps$) terms of the 2--loop heavy operator matrix elements (OMEs), 
\cite{UnPolOeps,KrakProc},
contributing to the 3--loop heavy flavor Wilson coefficients via renormalization. In 
the present paper we also report on new results contributing to the 3--loop 
operator matrix elements and the progress towards a
full 3--loop calculation of moments of the heavy flavor Wilson
coefficients. As part of the calculation, the $T_F^2$--terms of the 
even moments $N=2 ...10$ of the ${\sf NNLO}$ non-singlet (${\sf NS}$) and 
pure-singlet (${\sf PS}$)
anomalous dimensions given in \cite{LARIN,MVVandim}, cf. also \cite{GRACEY} 
in the {\sf NS}-case, 
are confirmed in an independent calculation.
%
%
\section{Heavy flavor Wilson coefficients in the asymptotic limit}

\noindent
As outlined in Ref.~\cite{BUZA}, the heavy flavor Wilson 
coefficients\footnote{We consider extrinsic charm production only.}
in the limit $Q^2 \gg m^2$ are obtained as a convolution of the light 
flavor Wilson coefficients with the corresponding massive operator matrix 
elements of the flavor decomposed quarkonic and gluonic operators between
massless parton states. Here we consider the level of twist--2 operators.
The light Wilson coefficients are known up to three loops \cite{MVV} and carry all 
the process dependence, whereas the OMEs are universal, process--independent 
objects. 
Their logarithmic contributions in $m^2/\mu^2$, as well as all pole terms in 
$1/\varepsilon$, are completely determined by renormalization. The latter 
provide
checks on  the calculation. Here, the single pole terms contain the respective
contributions of the 3-loop anomalous dimensions. 
%
%
\section{Renormalization}

\noindent
For the calculation of the bare heavy flavor OMEs we use dimensional 
regularization
in $D=4+\ep$ dimensions and work in the $\overline{\rm{MS}}$--scheme, if 
not stated otherwise, 
using Feynman--gauge. The renormalization is performed in four steps.
For mass renormalization we use the on--shell scheme
\cite{MASS2}, whereas charge renormalization is done using the
$\overline{\rm{MS}}$ scheme. For the latter we make the requirement
that the heavy quark loop contributions to the gluon self--energy, 
$\Pi(p^2,m^2)$, are renormalized in such a way that
$\Pi(0,m^2)=0$, cf. \cite{BUZA,BBK1,UnPolOeps}.
The origin of the remaining divergences is twofold. The UV singularities
are renormalized via the operator $Z$--factors, whereas the collinear
singularities are removed via mass factorization through 
the transition functions $\Gamma$. 
We denote the completely unrenormalized OMEs by a double--hat, 
$\,\hat{\hspace*{-1mm} \hat{A}}$, and
those for which mass and coupling renormalization have already
been performed by a single hat. Operator renormalization
and mass factorization then proceeds via 
\begin{eqnarray}
 {\bf A}&=&{\bf Z}^{-1} \hat{\bf A} {\bf \Gamma}^{-1}~. \label{GenRen}
\end{eqnarray}
Note that in the singlet case, Eq. (\ref{GenRen}) should be read as a
matrix equation, contrary to the ${\sf NS}$--case. The $Z$--factors
are related to the anomalous dimensions of the twist--$2$ operators
via
\begin{eqnarray}
 {\bf \gamma} = \mu \partial \ln {\bf Z(\mu)}/ \partial \mu~, 
\label{gamzet}
\end{eqnarray}
which allows  to express the $Z$--factors in terms of the anomalous
dimensions up to an arbitrary order in the strong coupling constant
$a_s := \alpha_s/(4\pi)$ (cf. \cite{UnPolOeps} up to $O(a_s^3)$). If all 
quarks 
were massless, the transition functions $\Gamma$ would be given 
by ${\bf \Gamma}={\bf Z}^{-1}$,
but since we are dealing with diagrams containing at least 
one heavy quark, this equation has to be modified in such 
a way that mass factorization is applied to those 
parts of the diagrams containing massless lines only. Finally, 
the ${\sf PS}$ and ${\sf NS}$ terms we are calculating are related by
\begin{eqnarray}
 Z^{\sf PS}_{qq}+Z^{\sf NS}_{qq}=Z_{qq} ~. \label{RelPSNS}
\end{eqnarray}
From Eqs. (\ref{GenRen},\ref{gamzet}) 
one thus can infer that for operator 
renormalization and mass factorization at $O(a_s^3)$,
the anomalous dimensions up to ${\sf NNLO}$, \cite{LARIN,MVVandim}, 
together with the $1$--loop heavy OMEs up to $O(\ep^2)$ and the 
$2$--loop heavy OMEs up to $O(\ep)$ are needed. The last two quantities 
enter since they multiply $Z-$ and $\Gamma$--factors containing 
poles in $\ep$. This has been worked out 
in some detail in Ref. \cite{UnPolOeps}, where we presented 
the $O(\ep)$ terms $\overline{a}_{Qg}^{(2)}$, 
$\overline{a}_{qq,Q}^{(2), {\sf NS}}$ and $\overline{a}_{Qq}^{(2){\sf PS}}$ 
in the unpolarized case. 
The term $\overline{a}_{gg,Q}^{(2)}$ was given in  \cite{KrakProc}.
The missing $2$--loop $O(\ep)$ term corresponds to the heavy OME
$A_{gq,Q}^{(2)}$ and was calculated for the first time 
in Ref. \cite{BUZA2}. It contributes through operator mixing to the
$T_F^2$--term of $A_{Qq}^{(3), {\sf PS}}$, which we consider in this paper. 
%
%
\section{\boldmath{$A_{gq,Q}^{(2)}$}}

\noindent
The term $A_{gq,Q}$ emerges for the first time at $O(a_s^2)$.
By applying Eq. (\ref{GenRen}),  one obtains at $O(a_s^2)$
the renormalized OME
\begin{eqnarray}
   A_{gq,Q}^{(2)} \hspace*{-1mm} = \hspace*{-1mm}
                     \hat{A}_{gq}^{(2)} \hspace*{-1mm}
                    +Z_{gq}^{-1,(2)} 
                    +\Bigl(
                           Z_{gg}^{-1,(1)}
                          +\hat{A}_{gg,Q}^{(1)}
                      \Bigr)\Gamma_{gq}^{-1,(1)}
                    . \label{AgqQ2Ren1}
\nonumber
\end{eqnarray}
Here, the term $\hat{A}_{gg,Q}^{(1)}$, cf. \cite{BUZA2},
enters through mixing. Note that since we consider only terms involving 
at least one heavy quark, we adopt the following definition for the 
anomalous dimensions
\begin{eqnarray}
 \hat{\gamma}\equiv \gamma(n_f+1)-\gamma(n_f)~ \label{hatpres}
\end{eqnarray}
in order to obtain the correct color projection. 
Now we can predict the structure of the unrenormalized result to be
\begin{eqnarray}
 \hat{A}_{gq,Q}^{(2)}
&=&\Bigl(\frac{m^2}{\mu^2}\Bigr)^{\ep}\Biggl[
                     \frac{2\beta_{0,Q}}{\ep^2}\gamma_{gq}^{(0)}
                    +\frac{\hat{\gamma}_{gq}^{(1)}}{2\ep}
                    +a_{gq,Q}^{(2)} \N\\ &&
                    +\ep\overline{a}_{gq,Q}^{(2)}
                     \Biggr]~, \label{AgqQ2Ren2}
\end{eqnarray}
where the ${\sf LO}$ and ${\sf NLO}$ anomalous dimensions 
are given by 
\begin{eqnarray}
   \gamma_{gq}^{(0)}&=&-4C_F\frac{N^2+N+2}{(N-1)N(N+1)}~, \label{ggq0}\\
   \hat{\gamma}_{gq}^{(1)}&=&C_FT_F\Biggl(
             -\frac{32}{3}\frac{N^2+N+2}{(N-1)N(N+1)}S_1 \N \\ &&
             +\frac{32}{9}\frac{8N^3+13N^2+27N+16}{(N-1)N(N+1)^2}
             \Biggr)~, \label{ggq1hat}
\end{eqnarray} 
and $\beta_{0,Q} = -4/3 T_F$~. The calculation in Mellin--space 
in terms of Feynman--parameters is straightforward, cf. \cite{BBK1}, 
and a representation in Euler--$\Gamma$ functions 
can  be obtained even to all orders in $\ep$ 
We 
reproduce the pole terms of Eq. (\ref{AgqQ2Ren2}) and obtain for the constant term in 
$\ep$ \begin{eqnarray}
   a_{gq,Q}^{(2)}&=&
          T_FC_F\Biggl\{
            \frac{4}{3}\frac{N^2+N+2}{(N-1)N(N+1)}
               \Bigl(S_2+S_1^2 \N \\ &&
               +2\zeta_2
               \Bigr)
           -\frac{8}{9}\frac{8N^3+13N^2+27N+16}
                            {(N-1)N(N+1)^2}S_1 \N\\ &&
           +\frac{8}{27}\frac{P_1}
                            {(N-1)N(N+1)^3}
                 \Biggr\}~,  \label{agqQ2} \\
   P_1&=&43N^4+105N^3+224N^2+230N+86~. \N
\end{eqnarray}
The last term has first been  calculated in \cite{BUZA2}, with which we 
agree. 
The $O(\ep)$ term is new and we obtain
\begin{eqnarray}
 \overline{a}_{gq,Q}^{(2)}&=&
          T_FC_F\Biggl\{
            \frac{2}{9}\frac{N^2+N+2}{(N-1)N(N+1)}
               \Bigl(-2S_3 \N\\ &&
             -3S_2S_1-S_1^3+4\zeta_3-6\zeta_2S_1\Bigr)\N\\ &&
           +\frac{2}{9}\frac{8N^3+13N^2+27N+16}
                            {(N-1)N(N+1)^2}
                \Bigl(2\zeta_2\N\\ &&
                +  S_2+S_1^2\Bigr)
           -\frac{4P_1S_1}
                             {27(N-1)N(N+1)^3}\N\\ &&
           +\frac{4P_2}
                             {81(N-1)N(N+1)^4}
                \Biggr\}~, \label{abgqQ2}\\
  P_2&=&248N^5+863N^4+1927N^3+2582N^2 \N\\ &&
         + 1820N+496~.  \N
\end{eqnarray}
Eqs. (\ref{ggq1hat}--\ref{abgqQ2})
are given in terms of harmonic sums, \cite{HS1,HS2}, the argument of which 
we have set equal to $N$. The representation in Mellin--space
allowed us to use various analytic and algebraic relations
between harmonic sums, \cite{ALGEBRA,STRUCT}, to obtain a more compact 
result.
Together with the result of Eq. (\ref{abgqQ2}), 
all $2$--loop $O(\ep)$ terms of the heavy OMEs in the unpolarized case
are known by now. As a last remark, note that we consider charm
quark contributions here, while for heavier quarks decoupling \cite{DECOUP}
has to be applied.
%
%
\section{\boldmath Fixed values of $N$ at three loops}

\noindent
The diagrams we need to calculate are of the $3$--loop self energy type
with the external particle being massless and on--shell and contain 
two inner scales. One is set by the mass of the heavy quark, the other 
by the Mellin--variable $N$ of the operator insertions emerging 
in the light--cone expansion. At this order, new operator vertices appear
with three and four gluonic lines, for which the Feynman--rules had not yet 
been derived. The necessary diagrams are generated using
{\sf QGRAF} \cite{Nogueira}. The number of diagrams contributing to 
$A^{(3)}_{Qg}$, e.g., is 1478 diagrams with one and 489 diagrams with two
quark loops. \\ \noindent
The steps for the calculation of these diagrams are 
the following: The diagrams are genuinely given as tensor integrals due to the
operators contracted with the light--cone vector $\Delta$, $\Delta^2=0$.  The
idea is, to first undo this contraction and to develop a projector, which,
applied to the tensor integrals, provides the results for the diagrams for the
specific (even) Mellin $N$ under consideration. So far, we implemented the
projector for the first 5 contributing Mellin moments, $N=2,...,10$,
where the color factors are calculated using \cite{COLORF}. A generalization 
to higher moments is straightforward, however, one quickly runs into 
computing time problems. The diagrams are
then translated into a form, which is suitable for the program {\tt MATAD}
\cite{MATAD}, which does the expansion in $\eps$ for the corresponding 
massive three--loop tadpole--type diagrams. We have implemented these
steps into a {\tt FORM}--program,  cf.~\cite{FORM},  and tested it 
against 
various two--loop results, including the result for
$\hat{A}_{gq,Q}^{(2)}$, Eq. (\ref{AgqQ2Ren2}), and found agreement.
%
%
\section{First results}

\noindent
 The first $3$--loop objects being investigated are the terms $\propto~T_F^2$ 
 of the OMEs $A_{qq,Q}^{\sf NS}$ and $A_{Qq}^{\sf PS}$. Note that in 
 the ${\sf NS}$--case, the $\pm~$ and $v$ terms do not differ. 
All diagrams 
contain two inner quark loops, where the quark to which the operator 
 insertion couples is heavy and the other one may be heavy or light. 
 The latter two cases can be distinguished by a factor $n_f$, denoting 
 the number of light flavors, in the result.
{\small
\begin{table*}[htb]
\caption{Mellin moments $2$ to $10$ for the $T_F^2$ terms 
of the $O(\ep^0)$ terms of the unrenormalized \newline
OME $\,\hat{\hspace*{-1mm} \hat{A}}_{Qq}^{(3),{\sf PS}}$
as obtained from {\tt MATAD}.}
\label{table:results1}
\newcommand{\m}{\hphantom{ }}
\newcommand{\cc}[1]{\multicolumn{1}{c}{#1}}
\renewcommand{\arraystretch}{2.0} 
\begin{tabular}{lllll}
\hline\hline
$\displaystyle \rm{N}$  & \cc{$\displaystyle 
    a_{Qq}^{(3),{\sf PS}}|_{T_F^2}$}   \\
\hline\hline
 $\displaystyle \rm{2}$  & 
 \m $\displaystyle 
        -\frac{36880}{2187}
        -\frac{736}{81}\zeta_2
        -\frac{4096}{81}\zeta_3
     +n_f\Bigl(
        -\frac{76408}{2187}
        -\frac{112}{81}\zeta_2
        +\frac{896}{81}\zeta_3
         \Bigr)
    $  \\
 $\displaystyle \rm{4}$  & 
 \m $\displaystyle 
        -\frac{2879939}{5467500}
        -\frac{1118}{2025}\zeta_2
        -\frac{15488}{2025}\zeta_3 
     +n_f\Biggl(
        -\frac{474827503}{109350000}
        -\frac{851}{20250}\zeta_2
        +\frac{3388}{2025}\zeta_3
         \Biggr)
    $  \\
 $\displaystyle \rm{6}$  & 
 \m $\displaystyle
         \frac{146092097}{1093955625}
        -\frac{7592}{99225}\zeta_2
        -\frac{61952}{19845}\zeta_3 
     +n_f\Biggl(
        -\frac{82616977}{45378900}
        -\frac{16778}{694575}\zeta_2
        +\frac{1936}{2835}\zeta_3
         \Biggr)
    $  \\
 $\displaystyle \rm{8}$  & 
 \m $\displaystyle
         \frac{48402207241}{272211166080}
        +\frac{1229}{142884}\zeta_2
        -\frac{43808}{25515}\zeta_3 
     +n_f\Biggl(
        -\frac{16194572439593}{15122842560000}
        -\frac{343781}{14288400}\zeta_2
        +\frac{1369}{3645}\zeta_3 
         \Biggr)
    $  \\
 $\displaystyle \rm{10}$  & 
 \m $\displaystyle
         \frac{430570223624411}{2780024890190625}
        +\frac{319072}{11026125}\zeta_2 
        -\frac{802816}{735075}\zeta_3 
    $  \\
 $\displaystyle \rm{ }$  & 
 \m $\displaystyle
     +n_f\Biggl(
        -\frac{454721266324013}{624087220246875}
        -\frac{547424}{24257475}\zeta_2
        +\frac{175616}{735075}\zeta_3
         \Biggr)
    $  \\
[3mm]
\hline\hline
\end{tabular}\\[2pt]
\end{table*}
\renewcommand{\arraystretch}{1.0} 
}
%
%
%
{\small
\begin{table*}[htb]
\caption{Mellin moments $2$ to $10$ 
of $\gamma_{\sf PS}^{(2)}|_{T_F^2}$ 
as obtained from {\tt MATAD} and Eq. (\ref{PoleStruct2}).}
\label{table:results2}
\newcommand{\m}{\hphantom{ }}
\newcommand{\cc}[1]{\multicolumn{1}{c}{#1}}
\renewcommand{\arraystretch}{2.0} 
\begin{tabular}{lccccc}
\hline\hline
$\displaystyle \rm{N}$  
   & \cc{$\displaystyle 2$} 
   & \cc{$\displaystyle 4$} 
   & \cc{$\displaystyle 6$} 
   & \cc{$\displaystyle 8$} 
   & \cc{$\displaystyle 10$}  \\
\hline\hline
 $\displaystyle \gamma_{\sf PS}^{(2)}|_{T_F^2}$
   & $\displaystyle  -\frac{5024}{243}
     $ 
   & $\displaystyle  -\frac{618673}{151875}
     $ 
   & $\displaystyle  -\frac{126223052}{72930375}
     $
   & $\displaystyle  -\frac{13131081443}{13502538000}
     $
   & $\displaystyle  -\frac{265847305072}{420260754375}
     $  \\
[3mm]
\hline\hline
\end{tabular}\\[2pt]
\end{table*}
\renewcommand{\arraystretch}{1.0} 
}
%
%
\subsection{\boldmath $A_{Qq}^{(3),{\sf PS}}$}
 
\noindent
From Eq. (\ref{GenRen}), we obtain the pole structure 
 of the completely unrenormalized ${\sf PS}$ OME. Considering 
 only the $T_F^2$ terms, one finds
\begin{eqnarray}
&& \hspace*{-7mm}  
\,\hat{\hspace*{-1mm} \hat{A}}_{Qq}^{(3),{\sf PS}}\Bigg|_{T_F^2}=
   \Bigl(\frac{m^2}{\mu^2}\Bigr)^{3\ep/2}\Biggl\{
     2\frac{n_f+4}{3\ep^3}\beta_{0,Q}\hat{\gamma}_{qg}^{(0)}\gamma_{gq}^{(0)}
   \N \\ 
&& \hspace*{-7mm}  
      +\frac{1}{\ep^2}       \Bigl(
                           \frac{2-n_f}{6}\hat{\gamma}_{qg}^{(0)}
                                      \hat{\gamma}_{gq}^{(1)}
                      -(n_f+1)\frac{4}{3}\beta_{0,Q}\hat{\gamma}_{\sf PS}^{(1)}
                              \Bigr) \N\\ 
&& \hspace*{-7mm}  
       +\frac{1}{\ep}         \Bigl(
                     \frac{n_f+1}{3}\hat{\overline{\gamma}}_{\sf PS}^{(2)}
                    -4(n_f+1)\beta_{0,Q}a_{Qq}^{(2),{\sf PS}} \\ 
&& \hspace*{-7mm}  
                    -n_f\frac{\zeta_2 \beta_{0,Q}}{4}\hat{\gamma}_{qg}^{(0)}
                                \gamma_{gq}^{(0)}
                    +\hat{\gamma}_{qg}^{(0)}a_{gq,Q}^{(2)}
                              \Bigr) 
\hspace*{-1mm}        
+a_{Qq}^{(3),{\sf PS}}\Bigg|_{T_F^2}
                                         \Biggr\}, \label{PoleStruct2}
\nonumber
 \end{eqnarray}
where we have written the $n_f$ dependence explicitly 
 and with $\hat{\overline{\gamma}}_{\sf PS}^{(2)}$
 being the term $\propto~n_f^2$ of the ${\sf NNLO}$ anomalous dimension 
 $\gamma_{\sf PS}^{(2)}$. It is not possible 
 to factor out $n_f+1$, not even in the triple pole term. This is due 
 to the interplay of the prescription for coupling constant 
 renormalization we have adopted, cf. \cite{BUZA,UnPolOeps}, and the fact 
 that the transition functions $\Gamma$ apply to sub graphs
 containing massless lines only. 
 We have calculated the above term
 using {\tt MATAD} for $N=2,...,10$ and all pole terms agree 
 with Eq. (\ref{PoleStruct2}). In Table \ref{table:results1},
 we show the constant terms in $\ep$ we have obtained. 
%
%
%
%
 
Using Eqs. (\ref{agqQ2},\ref{PoleStruct2}), 
 one can obtain moments for the $3$--loop anomalous dimension 
 $\gamma_{\sf PS}^{(2)}|_{T_F^2}$, which we show in Table 
 \ref{table:results2}. These latter results agree with 
 the results from Refs. \cite{MVVandim}.
 Here one 
 has to make the replacement  
{\small
\begin{table*}[htb]
\caption{Mellin moments $2$ to $10$ for the $T_F^2$ terms 
of the $O(\ep^0)$ terms of the unrenormalized OME \newline
$\,\hat{\hspace*{-1mm} \hat{A}}_{qq,Q}^{(3),{\sf NS}}$
as obtained from {\tt MATAD}.}
\label{table:results3}
\newcommand{\m}{\hphantom{ }}
\newcommand{\cc}[1]{\multicolumn{1}{c}{#1}}
\renewcommand{\arraystretch}{2.0} 
\begin{tabular}{lllll}
\hline\hline
\hspace*{-3mm}
$\displaystyle \rm{N}$  & \cc{$\displaystyle 
    a_{qq,Q}^{(3),{\sf NS}}|_{T_F^2}$}   \\
\hline\hline
\fs
\hspace*{-3mm}
 $\displaystyle \rm{2}$  & \fs 
 \m $\displaystyle 
\hspace*{-8mm}
          - \frac{28736}{2187}
          - \frac{512}{81}\zeta_2 
          - \frac{2048}{81}\zeta_3
          +n_f\Bigl(
               -\frac{100096}{2187}
               -\frac{256}{81}\zeta_2
               +\frac{896}{81}\zeta_3
              \Bigr) 
    $  \\
\hspace*{-3mm} \fs
 $\displaystyle \rm{4}$  & 
 \m 
\hspace*{-8mm}
$\displaystyle 
          - \frac{151928299}{5467500}
          - \frac{26542}{2025}\zeta_2
          - \frac{20096}{405}\zeta_3
          +n_f\Bigl(
               -\frac{1006358899}{10935000}
               -\frac{13271}{2025}\zeta_2
               +\frac{8792}{405}\zeta_3
              \Bigr) 
    $  \\
\hspace*{-3mm}
 $\displaystyle \rm{6}$  & 
 \m \hspace*{-8mm}
$\displaystyle
          - \frac{26884517771}{729303750}
          - \frac{1712476}{99225}\zeta_2
          - \frac{181504}{2835}\zeta_3
          +n_f\Bigl(
               -\frac{524427335513}{4375822500}
               -\frac{856238}{99225}\zeta_2
               +\frac{11344}{405}\zeta_3
             \Bigr)
    $  \\
\hspace*{-3mm}
 $\displaystyle \rm{8}$  & 
 \m \hspace*{-8mm}
$\displaystyle
          - \frac{740566685766263}{17013197880000}
          - \frac{36241943}{1786050}\zeta_2
          - \frac{632512}{8505}\zeta_3
              +n_f\Bigl(
               -\frac{
4763338626853463}{34026395760000}
               -\frac{36241943}{3572100}\zeta_2
               +\frac{39532}{1215}\zeta_3
              \Bigr)
    $  \\
\hspace*{-3mm}
 $\displaystyle \rm{10}$  & 
 \m \hspace*{-8mm}
$\displaystyle
          -\frac{6080478350275977191}{124545115080540000}
          -\frac{2451995507}{108056025}\zeta_2
          -\frac{1543040}{18711}\zeta_3
  $  \\
 $\displaystyle\phantom{ \rm{10}}$  & 
 \m $\displaystyle
          +n_f\Bigl(
          -\frac{38817494524177585991}{249090230161080000}
          -\frac{2451995507}{216112050}\zeta_2
          +\frac{96440}{2673}\zeta_3
              \Bigr) 
    $  \\
[3mm]
\hline\hline
\end{tabular}\\[2pt]
\end{table*}
\renewcommand{\arraystretch}{1.0} 
}
%
%
%
%
%
%
{\small
\begin{table*}[htb]
\caption{Mellin moments $2$ to $10$ of
$\gamma_{\sf NS}^{(2)}|_{T_F^2}$ as
obtained from {\tt MATAD} and Eq. (\ref{GenRen}).}
\label{table:results4}
\newcommand{\m}{\hphantom{ }}
\newcommand{\cc}[1]{\multicolumn{1}{c}{#1}}
\renewcommand{\arraystretch}{2.0} 
\begin{tabular}{lccccc}
\hline\hline
$\displaystyle \rm{N}$  
   & \cc{$\displaystyle 2$} 
   & \cc{$\displaystyle 4$} 
   & \cc{$\displaystyle 6$} 
   & \cc{$\displaystyle 8$} 
   & \cc{$\displaystyle 10$}  \\
\hline\hline
 $\displaystyle \gamma_{\sf NS}^{(2)}|_{T_F^2}$
   & $\displaystyle  
         -\frac{1792}{243}
     $ 
   & $\displaystyle  
       -\frac{384277}{30375}
     $ 
   & $\displaystyle  
      -\frac{160695142}{10418625}
     $
   & $\displaystyle  
      -\frac{38920977797}{2250423000}
     $
   & $\displaystyle  
         -\frac{27995901056887}{1497656506500}
     $  \\
[3mm]
\hline\hline
\end{tabular}\\[2pt]
\end{table*}
\renewcommand{\arraystretch}{1.0} 
}
$n_f\rightarrow~n_f(2T_F)$, with 
 $T_F=1/2$, and multiply with $2$, to account for the different 
 convention for the $Z$--factors we adopted, see Eq.~(\ref{gamzet}). 
 Note that in Tables (\ref{table:results1}--\ref{table:results4}), 
 there is an overall factor $C_FT_F^2$, which we do not 
 show explicitly.  \\ \noindent
 As an example consider the renormalized result for the second moment. 
 Applying Eq. (\ref{GenRen}), we obtain
  \begin{eqnarray}
&& \hspace*{-7mm}  
A_{Qq}^{(3),{\sf PS}}\Bigg|_{N=2, T_F^2}=
           C_FT_F^2\Biggl\{
           -\frac{128}{81}
              \ln^3 \Bigl(\frac{m^2}{\mu^2}\Bigr)            
\N\\ 
&& \hspace*{-7mm}  
-\frac{32}{27}
              \ln^2 \Bigl(\frac{m^2}{\mu^2}\Bigr)
           -\frac{5344}{243} 
              \ln \Bigl(\frac{m^2}{\mu^2}\Bigr)
           +\frac{53144}{2187} 
\N\\
&& \hspace*{-7mm}  
           -\frac{3584}{81}\zeta_3 
     +n_f\Biggl(
            -\frac{128}{81}
              \ln^3 \Bigl(\frac{m^2}{\mu^2}\Bigr) 
            +\frac{32}{27}
              \ln^2 \Bigl(\frac{m^2}{\mu^2}\Bigr)
\N\\ && \hspace*{-7mm}  
            -\frac{5104}{243} 
              \ln \Bigl(\frac{m^2}{\mu^2}\Bigr) 
            -\frac{34312}{2187}
            +\frac{1024}{81}\zeta_3
         \Biggr)
           \Biggr\}~. \label{RenResN2PS}
\end{eqnarray}
 As in Eq. (\ref{RenResN2PS}), we observe for all moments in 
 the ${\sf NS}$ and ${\sf PS}$ case that the 
 terms $\propto~\zeta_2$ disappear after renormalization, since the 
corresponding terms in  the light flavor Wilson coefficients do not 
contain even $\zeta$-values. This 
 provides us with a further check on our calculation, since it 
 is a general observation made in many $D=4$ calculations.
%
%
%
%
%
%
%
%
\subsection{\boldmath $A_{qq,Q}^{(3),{\sf NS}}$}

\noindent
For the $T_F^2$--terms of the heavy OME $A_{qq,Q}^{(3), \sf NS}$, 
a formula similar to Eq. (\ref{PoleStruct2}) can be derived, which we do 
not show here. 
Using again {\tt MATAD}, we have calculated the first $5$ non-vanishing
moments of the completely unrenormalized expression. 
We list the constant terms in $\ep$, $a_{qq,Q}^{(3),\sf NS}$, 
of our results in Table \ref{table:results3}. The pole terms
we obtain agree with what one expects from Eq. (\ref{GenRen})
and after renormalization, we again observe that there are no $\zeta_2$'s
left anymore. The values for the moments of the 
terms $\propto~T_F^2$ in $\gamma_{\sf NS}^{(2)}$ we thus obtain
are shown in Table \ref{table:results4}. These values agree with 
those in Refs. \cite{LARIN,GRACEY}. 
%
%
%
\section{Conclusions and Outlook}

\noindent
We calculated the last missing O($\eps$) contribution to 
the unpolarized heavy OMEs for general Mellin variable
$N$ at $O(a_s^2)$, needed for the renormalization at $O(a_s^3)$.
Furthermore, we
installed a program chain to calculate the corresponding 3--loop
diagrams to $O(a_s^3)$ using {\tt MATAD}. This chain is now
existing and we expect first complete results in the near future.
As a first step, we presented moments of the terms 
$\propto~T_F^2$ of the heavy OMEs
$\hat{A}_{qq,Q}^{(3),{\sf NS}}$ and $\hat{A}_{Qq}^{(3),{\sf PS}}$, 
for which we found agreement with the general pole structure expected
from renormalization. This provides us with a good check on the method 
we apply for our calculation. For the calculation of high moments we will
apply {\tt TFORM}, \cite{TFORM}, in the future.


\vspace{5mm}\noindent
{\bf Acknowledgments.}~~We would like to thank M.~Steinhauser 
and J. Vermaseren for useful discussions and M. Steinhauser for a
{\tt FORM 3.0} compatible form of the code {\tt MATAD}. 


\end{document}